\documentclass{aa}
\usepackage{graphics}
\usepackage{graphics}

\newcommand{\ep}{${{E_{\mathrm p}}}$}

\newcommand{\rxg} {$R_{\mathrm x/\gamma}$}

\begin{document}

\title{On the nature of X-Ray Flashes}

\author{C. Barraud\inst{1}, 
        F. Daigne\inst{2,3}, 
        R. Mochkovitch\inst{2}, and 
        J.L. Atteia\inst{1}}

   \offprints{C. Barraud}

\institute{$^1$Laboratoire d'Astrophysique de Toulouse-Tarbes, OMP, 31400 Toulouse, France 
\\
           $^2$Institut d'Astrophysique de Paris, 
98 bis, boulevard Arago, 75014 Paris, France\\
$^3$Universit\'e Pierre et Marie Curie -- Paris VI, 4, Place Jussieu,
75005 Paris, France }

   \date{}

   \abstract{We discuss the origin of X-Ray Flashes (XRFs), a recently
discovered class of Gamma-Ray Bursts (GRBs). Using a simplified model
for internal shocks we check if XRFs can be intrinsically soft due to 
some specific values
of the parameters describing the relativistic outflow emerging from the 
central engine. We generate a large number of synthetic events and find
that XRFs are obtained when the contrast 
$\Gamma_{\rm max}/\Gamma_{\rm min}$ of the Lorentz factor distribution
is small while the average Lorentz factor $\bar \Gamma$ is large. A few 
XRFs may be GRBs at large redshifts but we exclude this possibility for 
the bulk of the population. If outflows with a small contrast 
are commonly produced, even a large population of XRFs could be 
explained. If conversely the Lorentz factor distribution 
within the wind is broad, one should then rely on extrinsic causes, 
such as viewing angle
effects or high redshift.

      }

\maketitle

\section{Introduction}

An intriguing discovery in recent years is the existence 
of a population of soft gamma-ray bursts (GRBs) with little or no 
emission above 50 keV (\cite{heis01}, Kippen et al. 2001, \cite{barr03}). 
These events, which have been called X-Ray Flashes (XRFs), share a number 
of cha\-rac\-te\-ristics with the classical GRBs 
(long duration, non-thermal spectra...),
and there is now general consensus on the fact that XRFs represent an extension 
at low e\-ner\-gies of the GRB population. 
In this context it is natu\-ral to verify whether the models de\-ve\-lo\-ped to explain the prompt emission
of GRBs can also explain a po\-pulation of soft bursts like the XRFs.
This is a complex issue because XRFs can be explained 
either by {\it extrinsic factors} (e.g. viewing angle, redshift) 
or {\it intrinsic factors} 
(e.g. Lorentz factor, energy deposition...).
An overview of the factors that could give rise to soft GRBs appeared in  
Zhang \& Meszaros (2002).
Recently several authors have discussed in detail the effects of 
the viewing angle on the softness of GRBs (e.g. \cite{yama02},
\cite{zhan04}, \cite{lamb04}).

In this paper we concentrate on the impact of intrinsic parameters and
we specifically address the following question:
can the internal shocks model, which successfully explains many properties 
of the GRB prompt emission, also explain XRFs without calling
upon a particular set of extrinsic factors ?
Our work is based on an ana\-ly\-ti\-cal model that captures the essential 
physics of internal shocks. 
We demonstrate that internal shocks can produce XRFs quite naturally 
and we discuss the conditions required for this to happen.

The paper is organized as follows.
Section \ref{observ} summarizes our current knowledge of XRFs. 
Section \ref{model} introduces our analytical model of internal shocks.
Section \ref{results} presents the results of the simulation of 
a large number of GRBs and discusses the conditions required
to produce XRFs.
Section \ref{conclusion} summarizes our results and presents our conclusions.

\section{Observational properties of XRFs} \label{observ}
\subsection{Gamma-ray bursts and X-ray flashes}

In 2001, Heise et al. reported the discovery 
of XRFs, short transients detected by the Wide Field Cameras 
of BeppoSAX in the range [2-26 keV] but not seen above 40 keV by the GRBM on-board
the same spacecraft
(see \cite{boel97} for a description of the BeppoSAX mission).
In order to cla\-ri\-fy the relationship between  
XRFs and GRBs,
Heise et al. (2001) compared the pro\-per\-ties 
(duration and spectral hardness) of 9 XRFs
with the 
X-ray counterparts of 16 GRBs also detected in the Wide Field Cameras of BeppoSAX. 
They concluded that ``the statistical pro\-per\-ties of XRFs display in all aspects
a na\-tu\-ral extension of the properties of GRBs''.
\cite{kipp01} (2001) a\-na\-ly\-zed the spectra of 9 XRFs simultaneously detected by BeppoSAX/WFC
and BATSE and found that XRF spectra, like those of GRBs, are well fitted 
by the
so-called Band function consisting of two smoothly connected power laws
(\cite{band93}). Defining $x=E/E_p$, where $E_p$ is the peak of the 
$E^2N(E)$ spectrum we have
\begin{equation}
N(E)\propto
\left\lbrace\begin{array}{l}
x^{\alpha}\,{\rm exp}(-(2+\alpha)\,x)\ \ \ \ \ {\rm if}\ x\le {\alpha-\beta
\over 2+\alpha}, \\
x^{\beta}\left({\alpha-\beta\over 2+\alpha}\right)^{\alpha-\beta}
{\rm exp}(\beta-\alpha)\ \ \ \ \ {\rm otherwise}
\end{array}\right.
\end{equation}
$\alpha$ and $\beta$ being the two slopes respectively 
at low and high energy.
The extended e\-ner\-gy range of HETE-2 (from 2 keV to 400 keV) 
has allowed to further a\-na\-lyze the relationship between XRFs and GRBs
(\cite{barr03}, \cite{barr04a}, \cite{saka04}, \cite{lamb04}).
It appears from these studies that XRFs are a continuation of the 
class of long GRBs at low energy. The distribution of their duration is 
indeed consistent with long GRBs. 
Heise et al. (2001) showed that the $t_{90}$ of 17 XRFs detected by BeppoSAX 
ranges from 10 s to 200 s and is comparable to the distribution 
of $t_{90}$ for the 36 GRBs studi\-ed.
The distribution of their spectral parameter $\alpha$ is also in agreement 
with the observed distribution for long GRBs. 
Kippen et al. (2001) showed that the distribution of $\alpha$ in the sample of XRFs and 
GRBs they 
stu\-died is consistent with what is found for bright BATSE bursts. 
Moreover, Barraud et al. (2004a) found values of 
$\alpha$ within the range predicted by 
the synchrotron shock models ($-3/2 \le \alpha \le -2/3$), 
whatever the value of $E_o$ ($E_o$ being related to the peak energy $E_p$ 
by the relation $E_o=E_p/(2+\alpha)$).
The distribution of the spectral parameter $\beta$ is also comparable 
to the distribution of $\beta$ for long bursts with a mean value of $-2.5$.
XRFs extend the well-known hardness-intensity correlation to 
soft, faint bursts (\cite{kipp01} 2001, \cite{barr03}).
Lamb et al. (2004) and Sakamoto et al. (2004) showed that XRFs 
also follow and extend the $E_{iso}$ - $E_p$ relation discovered 
by Amati et al. (2002).
It is therefore now generally accepted that the XRFs, XRRs (X-Ray Rich GRBs)
and classical long GRBs form a continuum, and that they share a common origin.   
\vskip 0.3cm
Using GRBs detected by BATSE, Preece et al. (2000) found a distribution 
of the peak energy 
that is narrow and centered around 200 keV. 
With the discovery of XRFs, Heise et al. (2001) and Kippen
et al. (2001) have shown that this distribution is broader 
than previously thought and that it is extended towards low energies, 
down to a few keV. 

One of the first explanations proposed for the XRFs was that 
they could be GRBs observed at very high redshifts (Heise et al. 2001). 
This hypothesis was however discarded by Barraud et al. (2003), in view
of the similar duration distributions of XRFs and long GRBs. 
Additionally, the first upper limits and measured spectroscopic redshifts 
for XRFs contradicts the high redshift hypothesis with 
XRF 020903 at $z=0.25$ (\cite{sode04}), XRF 040701 at $z=0.215$
(Kelson et al. 2004), XRF 011030 at $z<3.5$ (Bloom et al. 2003),
XRF 020427 at $z<2.3$ (Amati et al. 2004)
and XRF 030723 at $z<2.1$ (Fynbo et al. 2004).

The remaining possibilities to explain XRFs are ($\it i$) GRBs 
with different intrinsic 
properties or ($\it ii$) standard GRBs viewed off-axis.
In this paper we consider option ($\it i$) and  
study if the internal shock model of GRBs
can also account for XRFs and we determine the conditions required 
to produce them.

\subsection{Defining an XRF}
While XRFs are best defined by their \ep , the photon e\-ner\-gy 
of the maximum of their
$\nu F_\nu$ spectrum, \ep\ is not always available for weak soft events.
Consequently, we prefer to use the ratio \rxg\ of the $2-30$  
to the $30-400$ keV fluences to classify bursts into XRFs or GRBs.
This fluence ratio is easier to compute and more robust than \ep , and it 
has been shown that it closely reflects the value of \ep\ when it can
be measured (\cite{barr04a}). 
Following Sakamoto et al. (2004) we consider XRFs those events with \rxg $ \geq 1$.
This de\-fi\-nition calls for the following remarks.
First, the separation between XRFs and GRBs is somewhat arbitrary since 
the present data do not show a bimodal distribution of \ep~(some authors 
call X-Ray Rich GRBs 
intermediate events with \rxg\ in the range 0.3 to 1.).
Second, the true fraction of XRFs depends on the definition of XRFs, but
even more on the biases that affect their detection. Measuring the distances
of a few XRFs could be a first step towards estimating their true fraction in a given
volume.
For instance, GRB 030329 (at a redshift of 0.168) 
would have been classified as an XRF at a redshift larger than $z=2$.

With the above definition, seven transients stu\-di\-ed by Heise et al. (2001)
and seven transients stu\-di\-ed by Barraud et al. (2003) are XRFs.
Some XRFs have been described in detail in the literature; they include 
XRF 020903 and XRF 030723 detected by HETE-2 (\cite{saka04}, Butler et al.
2004) and 
XRF 020427 detected by Beppo-SAX (Amati et al. 2004). Three events,
GRB 981226  (\cite{fron00}), GRB 990704 (\cite{fero01}) 
and GRB 000615 (Maiorano et al. 2004) classified as X-ray rich GRBs 
by the SAX team   are XRFs 
according to our definition.

\section{XRFs in the context of the internal shock model} \label{model}
\subsection{A toy model for internal shocks}
The basic assumption of the internal shock scenario is that the central
engine of GRBs is able to generate a re\-la\-tivistic wind with a 
highly non-uniform
distribution of the Lorentz factor (with a contrast 
$\Gamma_{\rm max}/\Gamma_{\rm min}$ reaching at least a factor of 2). The observed emission is then produced
when layers of different velocities collide within the wind, the dissipated
energy being radiated in the gamma-ray range by means of synchrotron 
shock emission (Rees \& Meszaros, 1994). 
The evolution of this re\-la\-ti\-vistic wind
can be followed with a  
hydrodynamical si\-mu\-la\-tion (Daigne \& Mochkovitch, 2000) but this
requires large amounts of 
com\-pu\-ting time which 
prevents one from considering a large number of cases and fully explore the parameter
space. These detailed calculations have nevertheless shown that a simplified 
approach where the wind is represented by many shells which interact
by direct collisions can also produce good results (Kobayashi et al. 1997;
Daigne \& Mochkovitch, 1998). The reason
for this success  is 
that the kinetic energy of the wind largely dominates over its internal 
energy so that pressure waves can be neglected in a first appro\-xi\-mation. 
Going a step further we have de\-ve\-loped for this study a toy model where 
internal shocks are limited to the collision of only two shells of equal 
mass
$m$. Obviously, we cannot obtain from this toy model any detailed 
information on the temporal profiles but we expect that the main 
features of the burst  
energetics will be preserved.    
Shell 2 (of Lorentz factor $\Gamma_2$) is gene\-ra\-ted a time $\tau$ 
after shell 1 (of Lorentz factor $\Gamma_1<\Gamma_2$). This time
interval (multiplied by $1+z$) represents an order of magnitude of the
observed burst duration, since in the
internal shock model time scales seen by the observer 
reflect the source variability.
The average power 
injected by the central engine into the wind in this two shell approximation is given by
\begin{equation}
{\dot E}={mc^2\over \tau}(\Gamma_1+\Gamma_2)={\dot M}{\bar \Gamma}c^2
\end{equation}
where ${\dot M}=2m/\tau$ and ${\bar \Gamma}=(\Gamma_1+\Gamma_2)/2$ are the
average mass loss rate and Lorentz factor. Shell 2 will catch up with shell 1
at the shock radius
\begin{equation}
r_{\rm s}=2c\tau\,{\Gamma_1^2\Gamma_2^2\over \Gamma_2^2-\Gamma_1^2}
\end{equation}
The two shells merge at $r_{\rm s}$ and the energy dissipated in the collision
is given by
\begin{equation}
E_{\rm diss}=mc^2(\Gamma_1+\Gamma_2-2 \Gamma_{\rm s})
\end{equation}
where $\Gamma_{\rm s}=\sqrt{\Gamma_1 \Gamma_2}$ is the Lorentz factor of the
shocked material in the merged shell. In order to produce a GRB this 
energy has to be radiated in the gamma-ray range with a 
characteristic broken 
power-law spectrum.
If the synchrotron process is responsible for the emission, the peak energy
(ma\-xi\-mum of $\nu F_{\nu}$) is
\begin{equation}
E_{\rm p}\sim E_{\rm syn}=C_{\rm syn} \Gamma_{\rm s} B  \Gamma_e^2
\end{equation}
where  
$B$ and $\Gamma_e$ are the post shock magnetic field and electron
Lorentz factor
and $C_{\rm syn}={3\over 4\pi}{eh\over m_e c}$. 
Assuming that a fraction $\alpha_e$ of the dissipated
energy is transferred to a fraction $\zeta$ of the electrons we get
\begin{equation}
\Gamma_e={\alpha_e\over \zeta}{m_p\over m_e}\,{\epsilon}
\end{equation}
where $\epsilon c^2$ 
is the dissipated energy per unit mass in the comoving frame.

Similarly, if a fraction $\alpha_B$ of the energy goes into a disordered 
magnetic field generated behind the shock 
\begin{equation}
B=(8\pi \alpha_B \rho \epsilon c^2)^{1/2}
\end{equation}
the peak energy can be written as
\begin{equation}
E_{\rm p}=C_{\rm p}\,\Gamma_{\rm s}\rho^x\epsilon^y
\end{equation}
where $\rho$ is the post shock density, $x=1/2$, $y=5/2$ and
\begin{equation} 
C_{\rm p}=C_{\rm syn}\,(8\pi\alpha_B c^2)^{1/2}\left({\alpha_e\over \zeta}
{m_p\over m_e}\right)^2\,.
\end{equation}
We have considered below the possibility that $x$ and $y$ 
can be different from 1/2 and 5/2 if for example
the equipartition parameters 
are not constant but vary with $\rho$ and/or $\epsilon$.
The possibility of non-constant equipartition parameters 
has been considered by Chevalier (2003) and used in 
afterglow modelling by Yost et al. (2003) who assumed that $\alpha_B$ 
varies with the shock Lorentz factor. 
For the prompt phase, Daigne \&
Mochkovitch (2003) have shown 
that
the condition $2x+y<1$ (which therefore excludes the 
standard values $x=1/2$ and $y=5/2$) is
often required to obtain 
good fits of the
temporal and spectral evolution of GRB pulses. 
\\

In our two shell approximation, the physical parameters of the 
shocked layer $r_{\rm s}$, $\Gamma_{\rm s}$, 
$\rho$ and $\epsilon$ can be related to the wind quantities  
${\dot E}$, $\tau$, ${\bar \Gamma}$ and the contrast of Lorentz factor
$\kappa=\Gamma_2/\Gamma_1$ in the following way
\begin{equation}
r_{\rm s}=8c\tau {\bar \Gamma}^2 {\kappa^2\over (\kappa^2-1)(\kappa+1)^2}
\end{equation}
\begin{equation}
\Gamma_{\rm s}={2{\bar \Gamma}\over \kappa^{1/2}+\kappa^{-1/2}}
\end{equation}
\begin{equation}
\rho\sim {{\dot M}\over 4\pi r_{\rm s}^2 {\bar \Gamma} c}\sim 
{{\dot E}\over 256\,\pi c^5 \tau^2 {\bar \Gamma}^6}\,
\left(\kappa^2-1\right)^2\left(1+{1\over \kappa}\right)^4
\end{equation}
\begin{equation}
\epsilon={1\over 2}\left(\kappa^{1/2}+\kappa^{-1/2}\right)-1
\end{equation}
Replacing $\Gamma_{\rm s}$, $\rho$ and $\epsilon$ by their
expressions in Eq.8 yields
\begin{equation}
E_{\rm p}\propto {{\dot E}^x\over \tau^{2x}}{\varphi_{xy}(\kappa)\over
{\bar \Gamma}^{6x-1}}
\end{equation}
where the function
\begin{equation}
\varphi_{xy}(\kappa)={\left[\left(\kappa^2-1\right)\left(1+1/\kappa\right)^2
\right]
^{2x}\left(\kappa^{1/2}+\kappa^{-1/2}-2\right)^y\over 
\kappa^{1/2}+\kappa^{-1/2}}
\end{equation}
has been represented in Fig.1 for $\kappa=1$ to 10 and three choices of $x$ 
and $y$.
\begin{figure}
\resizebox{\hsize}{!}{\includegraphics{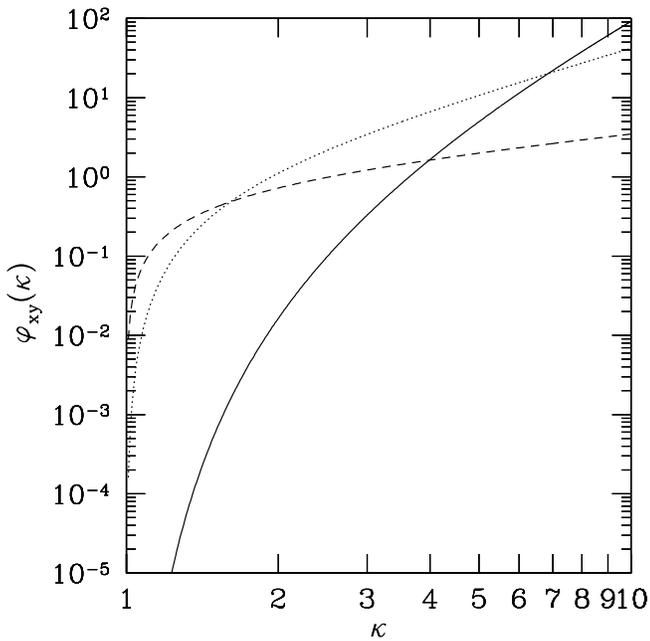}}
\caption{The function $\varphi_{xy}(\kappa)$ for $x=1/2$ and 
$y=5/2$ (full line)
$x=y=1/2$ (dotted line) and $x=y=1/4$ (dashed line). In the first case, the
large value of $y$ leads to a very steep dependence of $\varphi_{xy}$
on $\kappa$.}
\end{figure}

In spite of the simplicity of the two shell approximation, Eq.14 predicts an
anti-correlation between
duration and hardness as
observed in real bursts (Kouveliotou et al. 1993). Another important (and surprising) 
consequence of Eq.14 is that $E_{\rm p}$ is a decreasing function of 
${\bar \Gamma}$ as long as $x>1/6$. This can be understood from Eq.10
which shows that internal shocks occur closer to the source in a flow
with a low Lorentz factor, due to a large baryon load. If $x>1/6$ the reduced Lorentz factor cannot
compensate for the resulting increase of $\rho$ (Eq. 12) 
and a harder spectrum is
produced. To obtain softer bursts, ``clean fireballs'' 
(i.e. with a large ${\bar \Gamma}$) are required. 
This 
however only applies to cases where $\bar \Gamma$ remains sufficiently
high so that pair opa\-ci\-ty is unimportant at the source (Meszaros \&
Rees, 2000).
With pair creation
the situation becomes more complicated 
and has not been considered in this paper.
\subsection{Synthetic GRBs and XRFs}
The simplicity of the two shell
approximation allows us to construct large samples of synthetic
bursts to check if XRFs can be produced for some specific choice of the 
parameters.
A synthetic event is determined by the wind dynamics which
is fixed by the 
values of 
$\tau$,
$\dot E$, $\bar \Gamma$ and $\kappa$, the spectral parameters 
$C_{\rm p}$, $x$, $y$, $\alpha$ and
$\beta$ 
and the redshift $z$. 
For the spectral slopes we adopt $\alpha=-1$ and $\beta=-2.5$ 
which correspond to the average values obtained 
in spectral fits 
of bright long GRBs performed by Preece et al. (2000). We consider three different possible 
choices for
$x$ and $y$: ({\it{i}}) $x=1/2$, $y=5/2$, i.e. standard equipartion assumptions; 
({\it{ii}})
$x=y=1/2$, if for example the fraction of ac\-ce\-le\-ra\-ted electrons  
is proportional to $\epsilon$ so that $\Gamma_e$ remains 
approximately constant (Daigne \& Mochkovitch, 1998);
({\it{iii}}) $x=y=1/4$
which was used by Daigne \& Mochkovitch (2003) in their description 
of the temporal and spectral evolution of GRB pulses. These smaller va\-lues 
of $x$ and $y$ would correspond to a situation where the dependence of the
magnetic
field or/and electron Lorentz factor on the dissipated energy is much 
weaker than in the standard case. We believe that such a possibility cannot be  
excluded in view of the uncertainties in the microphysics
of the shocked material.

We generated a large number of events (from a few thousands to one million) 
by making assumptions on the distributions of the burst 
parameters, basing on observations or common hypothesis 
on the GRB physics and origin. The best constrained parameters 
are the redshift and the duration. 
If long GRBs (and XRFs) are related to the
explosive death of massive stars, their rate is 
directly proportional to the cosmic star formation rate $\psi_*$ 
and their distribution in redshift
can be deduced from $\psi_*(z)$ for which we have adopted the analytical 
expression
given by Porciani \& Madau (2001) with a maximum at $z\sim 1.5$ (their SFR 1). 
As shown by Bloom (2003) it remains presently impossible to decide 
between the three possible SFRs proposed by Porciani \& Madau (2001)
which can all be made compa\-ti\-ble with the present GRB redshift data
when corrected for high-redshift bias.
 
The distribution of the observed 
duration $t_{90}$ for long BATSE bursts is approximately log-normal 
with a ma\-xi\-mum at $t_{90}\sim20$ s. We have also adopted 
a log-normal distribution for $\tau$ with a 
maximum at $\tau_{\rm max}=10$ s and we checked a posteriori (see 
Sect. 4.3) that the resulting distribution of $\tau_{\rm obs}=(1+z)\tau$ 
for synthetic bursts   
agrees 
with that of $t_{90}$. 

The last four parameters $\bar \Gamma$, 
$\kappa$, $\dot E$ and $C_{\rm p}$ are much less constrained by observations
and we simply take for them 
uniform distributions between 100 and 500 
for $\bar \Gamma$, 0 and 1 for ${\rm Log}\,\kappa $, 51 and
53.4
for ${\rm Log}\,\dot E$ and 
${\rm Log}\,C_{\rm p}={\rm Log}\,C_{\rm p}^{100}\,\pm 0.5$ where 
$C_{\rm p}^{100}$
is the value of $C_{\rm p}$ which produces a typical 
burst with $E_{\rm p}=100$
keV if $\dot E=10^{52}$ erg.s$^{-1}$, $\bar \Gamma=300$, $\kappa=4$,
$\tau=5$ s
and $z=1$.  
The upper limit of 53.4 for ${\rm Log}\,\dot E$ 
has been estimated 
from the requirement that the synthetic ${\rm Log}\,N$-${\rm Log}\,P$ relation
agrees with the BATSE data (Stern et al. 2001). The comparison is shown in Fig.2 for 
$10^6$ 
synthetic events.

The assumptions of uniform distributions for $\bar \Gamma$, 
${\rm Log}\,\kappa $, ${\rm Log}\,\dot E$ and ${\rm Log}\,C_{\rm p}$
appear to be the simplest ones con\-si\-de\-ring our ignorance 
of the true distributions. 
The choice we have made also supposes that these quantities are independent,
which may be wrong. Thus, we cannot 
expect to obtain from our results any reliable estimate of the 
proportion of XRFs relative to GRBs, but we can identify  
the range of wind paramaters that favors the production of XRFs. We will
then know how the XRF/GRB ratio varies when
the distribution of these parameters differs from our 
initial simplest choice.     
   
\begin{figure}
\resizebox{\hsize}{!}{\includegraphics{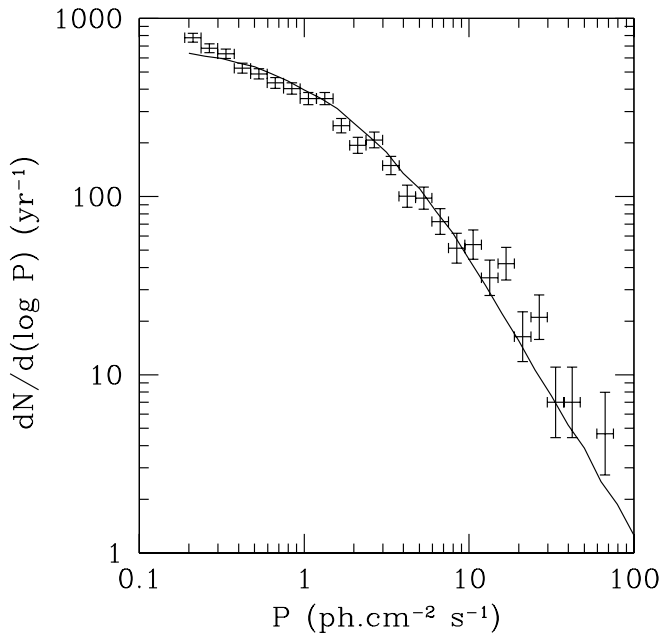}}
\caption{Differential peak flux distribution of BATSE bursts (Stern et al.
2001) compared to
a synthetic distribution ($10^6$ events) with 
$\left[{\rm Log}{\dot E}\right]_{\rm max}=53.4$.}
\end{figure}  
\section{Results} \label{results}
\subsection{$E_{\rm p}$ distribution and softness -- fluence relation}
We first obtained the $E_{\rm p}$ distribution of our synthetic bursts. The 
results are shown in Fig.3 for the three 
considered choices of $x$ and $y$.
The full line 
in Fig.3 represents the distribution for the whole sample while the
dotted and dashed
lines respectively correspond to the sub-groups of bursts 
which would have been detected
by BATSE and HETE 2. 
A threshold of 0.2 ph s$^{-1}$ cm$^{-2}$ in the 50 -- 300 keV 
energy range was assumed for BATSE while for HETE 2 we adopted 
1 ph s$^{-1}$ cm$^{-2}$ both for FREGATE (between 30 and 400 keV)
and the WXM (between 2 and 10 keV). 
These thresholds were estimated from the work of Band (2003) for the 
typical energy range of each ins\-trument. They are only indicative and
in practice also depend on the burst $E_{\rm p}$ and spectral
indices. This effect has not been included in our analysis. 
It should be rather moderate for FREGATE and the WXM 
due to the relative flatness of 
the sensitivity curves (Band, 2003).
In the case of BATSE it will contribute to decrease the already small
number of detected events at low energy.

For $x=1/2$ and $y=5/2$ and $x=y=1/2$
the 
distribution of $E_{\rm p}$ for BATSE bursts is wider than the observed one,
which is confined between 10 keV and 1 MeV (Preece et al. 2000). 
Conversely, the 
agreement is excellent for $x=y=1/4$ (since the value of 
$E_{\rm p}$ is then much less sensitive to the dispersion of the wind 
parameters). We have therefore adopted $x=y=1/4$
in the remainder of this paper since it appears that this choice of $x$
and $y$ gives the best results both for individual bursts
(Daigne and Mochkovitch, 2003) and statistically for a large popu\-la\-tion.

Figure 3 shows that BATSE misses most of the low $E_{\rm p}$ events
while HETE 2, which is 
less sensitive than BATSE in hard X-rays
but has a lower energy threshold, can detect at least part of them
down to $E_{\rm p}\sim $ a few keV as was the case for Beppo-SAX.       
\begin{figure}
\resizebox{\hsize}{!}{\includegraphics{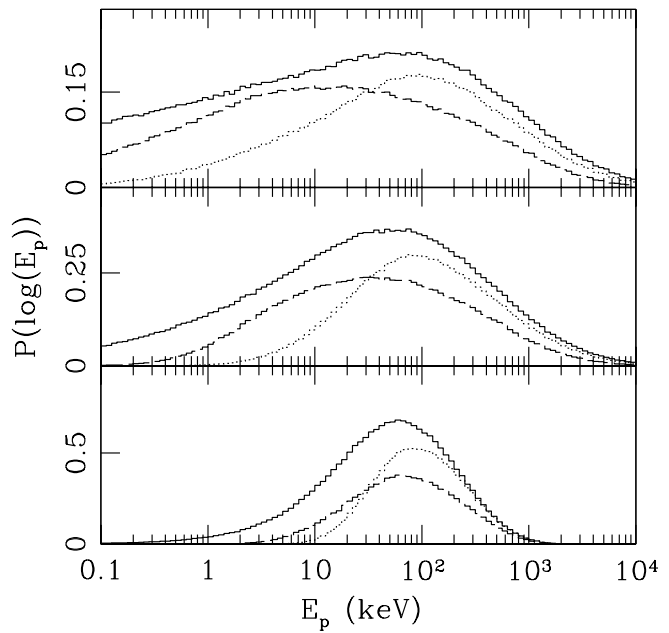}}
\caption{Distributions of $E_{\rm p}$ for bursts that can be detected 
respectively by
BATSE (dotted line) and by 
HETE 2 (dashed line). The
whole sample is represented by a full line; top panel: $x=1/2$, $y=5/2$; middle panel: $x=y=1/2$;
bottom panel $x=y=1/4$. The curves are normalized so that the integral for
the whole sample
$\int P({\rm log}E_{\rm p})\,d{\rm log}E_{\rm p}=1$.}
\end{figure}  
This is also illustrated in Fig.4
where we have represented the softness \rxg~as a function 
of the total (2 - 400 keV) fluence 
for a population of 1450 synthetic events which
would have been detected by HETE 2. The total number of events
produced was 3000, so that the detection fraction was about 1/2 (a 
smaller number of bursts was used in this case to avoid confusion in 
the figure).
The two limits of the 
softness at 0.075 and 4 respectively correspond 
to the hardest and softest bursts
for which the two bands 2-30 and 30-400 keV are both in the low or high 
energy part
of the spectrum. With the assumed values of $\alpha=-1$ and $\beta=-2.5$ 
the softness limits are 
simply given by 
\begin{equation}
R_{\mathrm x/\gamma}={\int_2^{30} dE\over \int_{30}^{400}dE}=0.0757
\end{equation}
for hard events and
\begin{equation}
R_{\mathrm x/\gamma}={\int_2^{30} E^{-1.5}dE\over \int_{30}^{400}
E^{-1.5}dE}=3.957
\end{equation}
for soft events.
The two horizontal dashed lines in Fig.4 separate the GRB, X-ray rich GRB and XRF 
domains. In agreement with Fig.3 it can be seen that the model gene\-ra\-tes 
a population of XRFs
that can be detected by HETE 2 (but would have mostly 
escaped detection by BATSE).   
Since, from the toy model, we know all the input parameters 
of these synthetic XRFs, 
we can identify the key ingredients necessary to
produce them. For this purpose we compare below the distributions of $z$, $\tau$,
$\dot E$, $\bar \Gamma$ and $\kappa$
between XRFs and standard GRBs. 
\begin{figure}
\resizebox{\hsize}{!}{\includegraphics{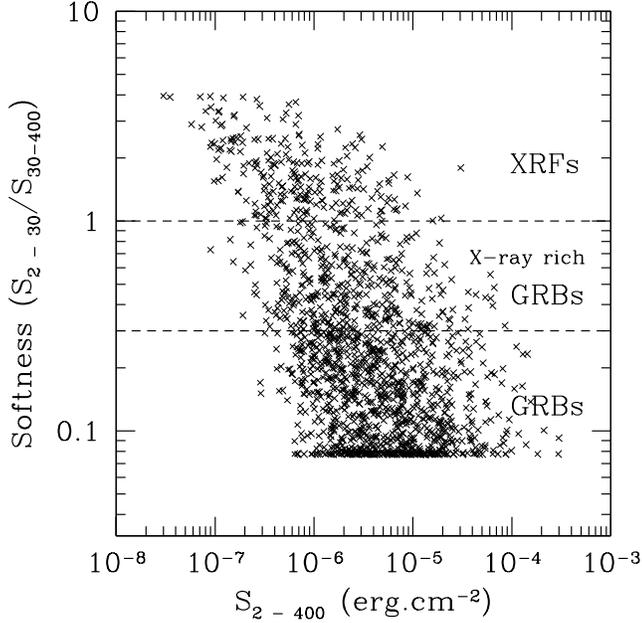}}
\caption{Softness versus total fluence for a population 
of 1450 synthetic events that can be detected by HETE 2. }
\end{figure}  
\begin{figure}
\resizebox{\hsize}{!}{\includegraphics{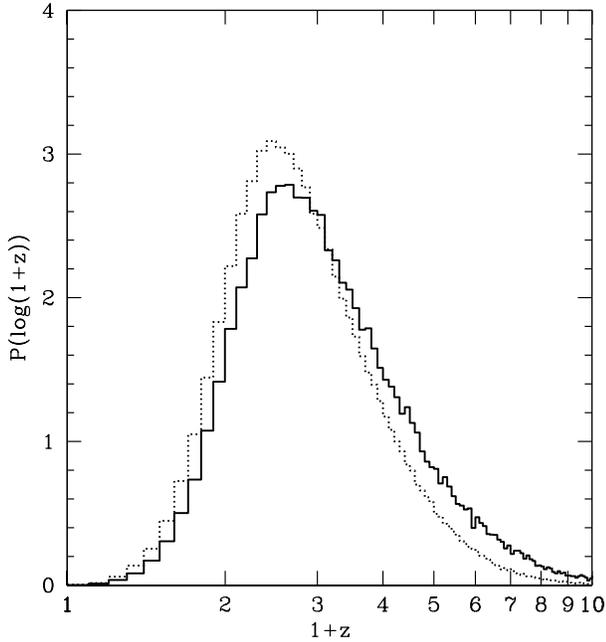}}
\caption{Redshift distribution of synthetic 
GRBs (dotted line) and XRFs (full line) obtained with our toy model.}
\end{figure} 
\subsection{Redshift distribution}
Figure 5 shows that the redshift distributions of synthetic GRBs 
(dotted line) and XRFs (full line) are very similar
(the adopted normalization is
$\int P({\rm log}(1+z))\,d{\rm log}(1+z)=1$). 
In the context of our simulation, XRFs are not standard GRBs observed at large $z$. 
This was already strongly suggested by (i) the duration distribution 
of the observed 
XRFs which 
is   comparable to that of long GRBs and (ii) the recent redshift
determinations (or upper limits) obtained for several XRFs (see Sect. 2.1).
Nevertheless, even at large redshift a bright GRB is still  
observable (Lamb \& Reichart, 2000) but can appear as an XRF 
so that the XRF/GRB ratio
is expected to increase with $z$. Fig.5 
shows that this is indeed the case: the XRF/GRB ratio at 
$z>5$ is more than 2 times larger than at $z=1$. However, the number 
of events at large $z$ 
is not sufficient to account for the 
bulk of the XRF population. 
We have checked that this was not a consequence of our specific choice for
$\psi_*(z)$ which is maximum at $z=1.5$. With SFR 2 of Porciani and Madau
(2001) which is nearly constant at $z>2$, the distributions of both
GRBs and XRFs remain similar, being only slightly shifted 
to larger $z$.
In any case, most events stay
confined between $z=1$ and 5 and the majority of XRFs are not 
GRBs at large $z$.  
     
Our synthetic XRFs are therefore intrinsically
soft due to some specific values of their relativistic wind parameters.
\subsection{Distribution of the wind parameters in GRBs and XRFs}
\begin{figure*}{!}
\begin{center}
\begin{tabular}{cc}
\resizebox{0.49\hsize}{!}{\includegraphics{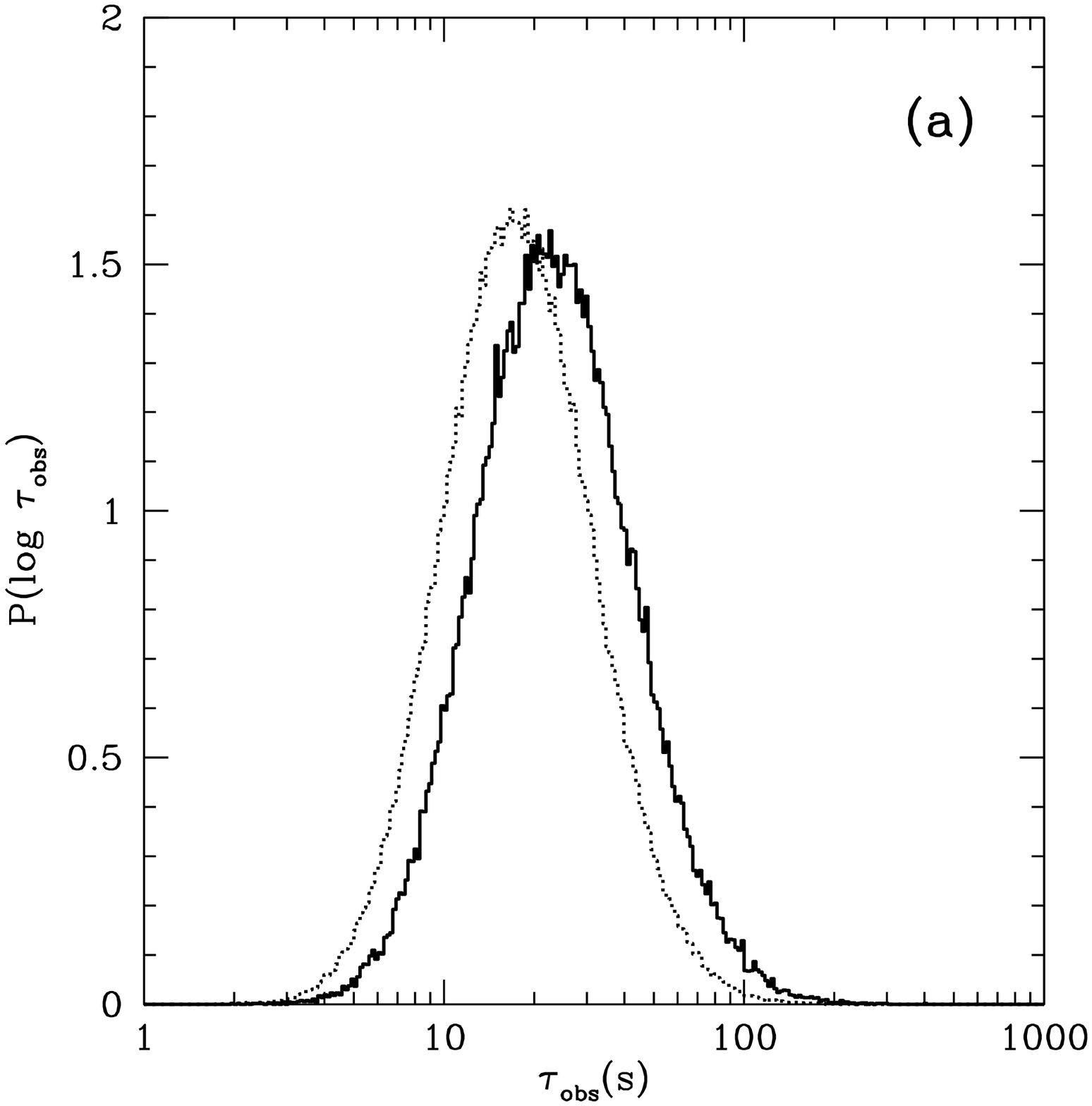}}&
\resizebox{0.49\hsize}{!}{\includegraphics{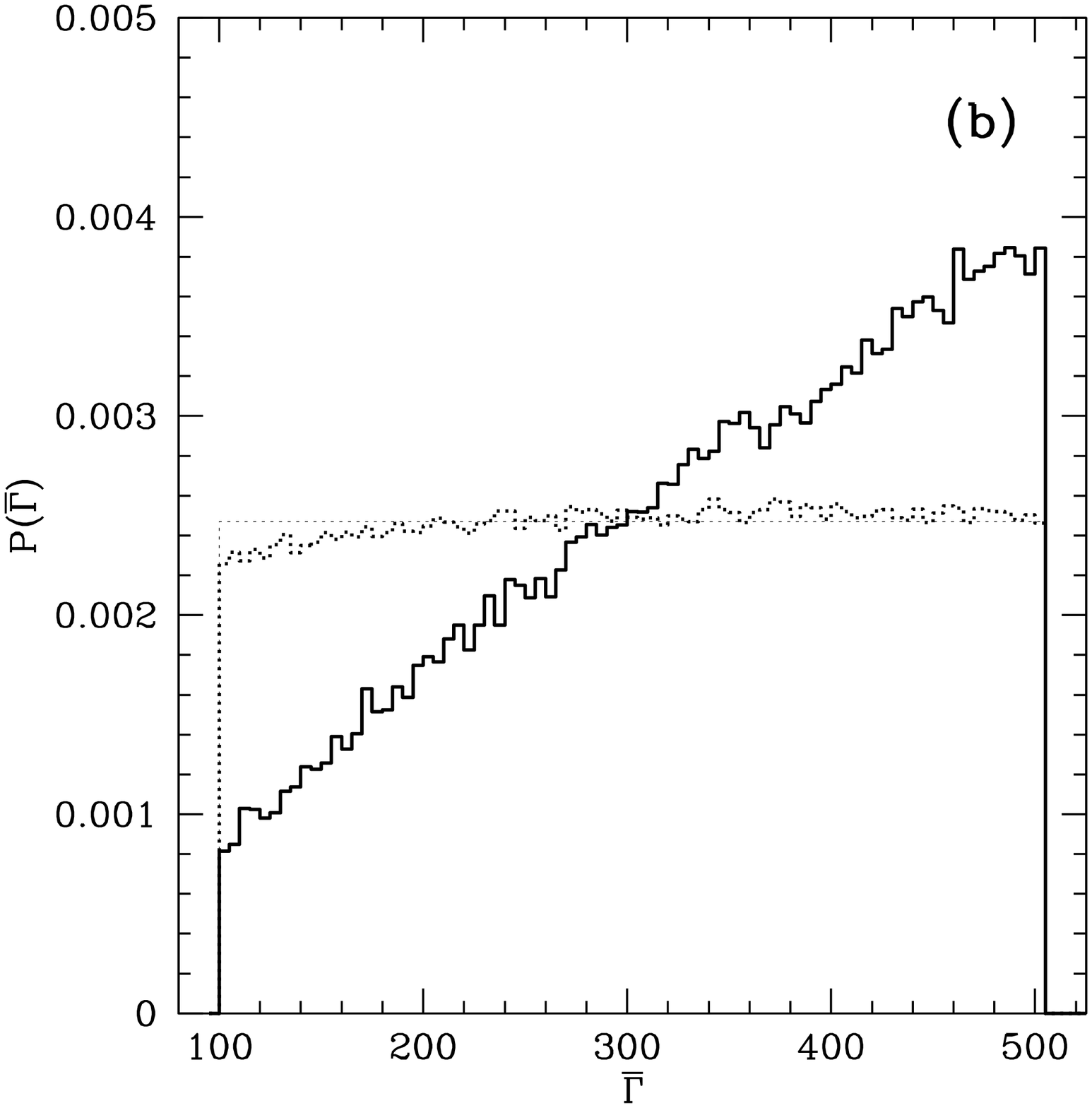}}\\
\resizebox{0.49\hsize}{!}{\includegraphics{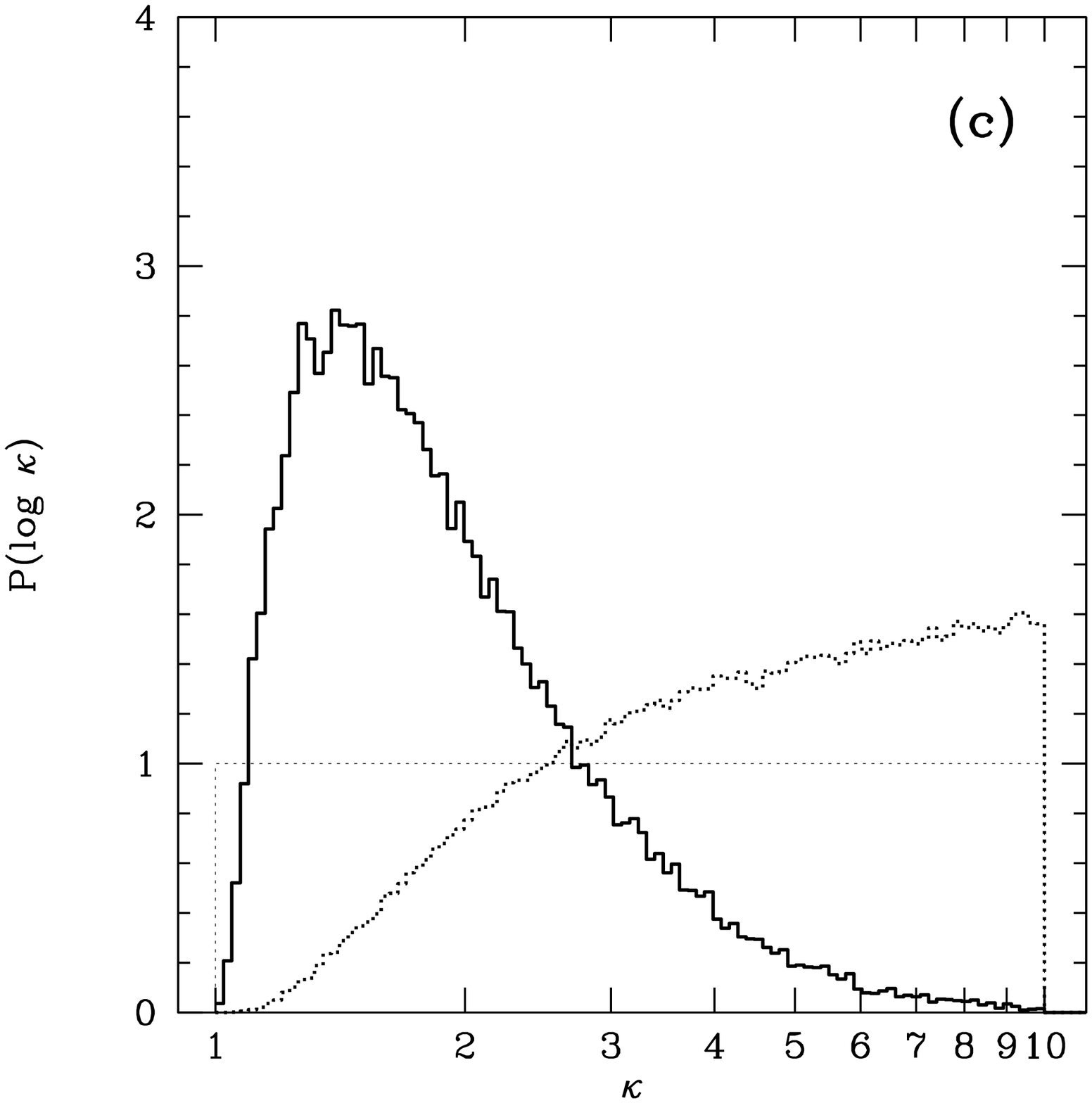}}& 
\resizebox{0.49\hsize}{!}{\includegraphics{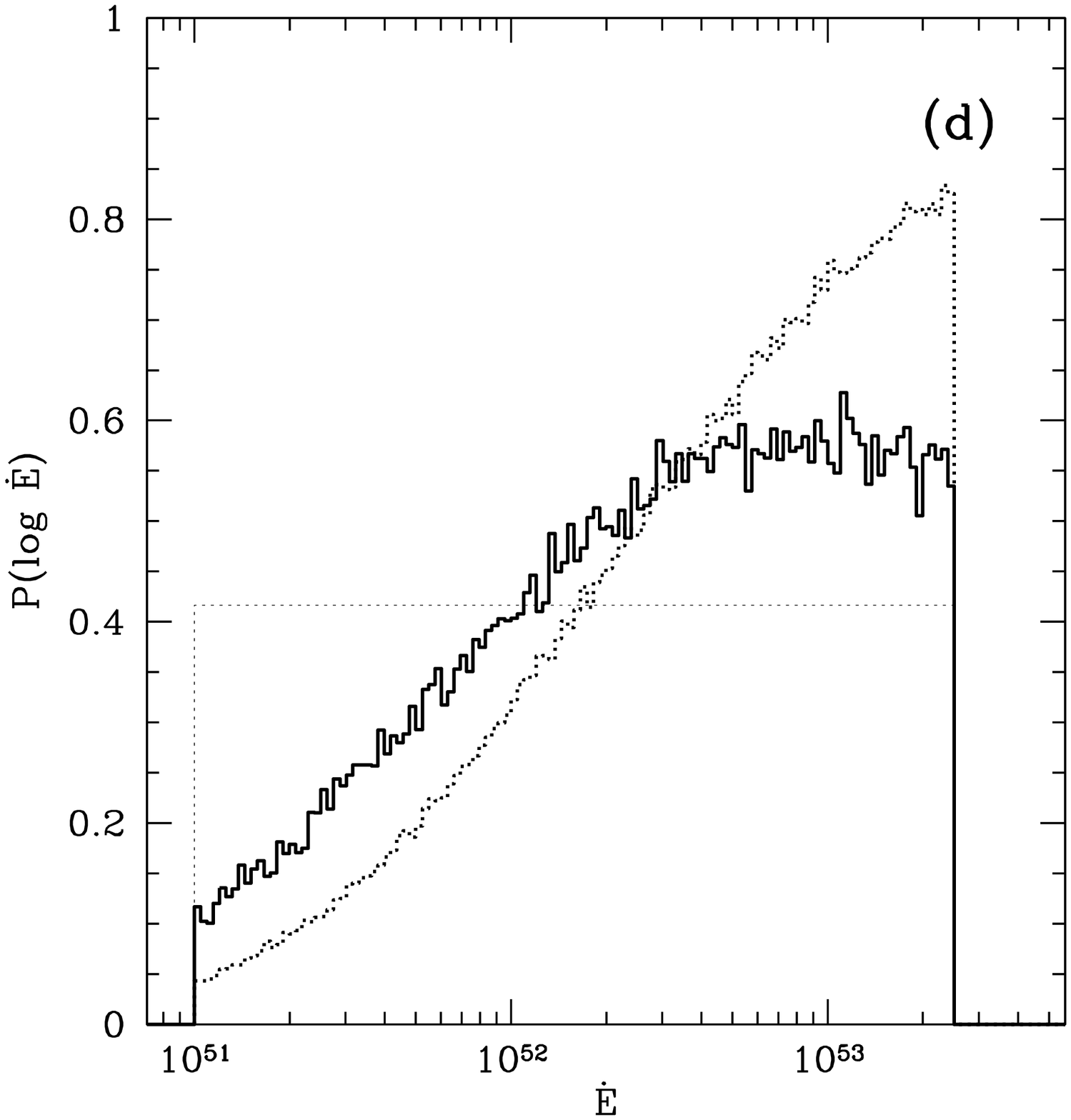}}\\
\end{tabular}
\end{center}
\caption{Distributions of $\tau_{\rm obs}=(1+z)\tau$ (a), $\bar \Gamma$ (b), 
$Log\,\kappa$ (c) and
$Log\,\dot E$ (d) for GRBs (dotted line) 
and XRFs (full
line). The thin dotted lines in (b, c, d) represent the uniform distributions 
used as input.
The adopted normalizations are 
$\int P(Log\,\tau_{\rm obs})\,dLog\,\tau_{\rm obs}=1$,\,
$\int_{100}^{500} P({\bar \Gamma})
d{\bar \Gamma}=1$,\, $\int_0^1 P(Log\,\kappa)\,dLog\,\kappa=1$ and
$\int_{51}^{53.4} P(Log\,\dot E)\,dLog\,\dot E=1$.}
\end{figure*}

We compare in Fig.6 the distribution of the four wind parameters ($\tau$,
$\bar \Gamma$,
$\kappa$, $\dot E$)  
in synthetic GRBs and XRFs.\vskip 0.2cm
\noindent{\it Observed duration:}\vskip 0.1cm
\noindent
The distribution of the observed duration $\tau (1+z)$ in GRBs and XRFs 
is represented in Fig.6a. 
As expected, it is in good agreement with the BATSE duration distribution for long GRBs. 
The average duration of XRFs is appro\-xi\-mately 50\% longer. 
This is not a consequence 
of a larger redshift but of a preferred longer intrinsic duration resul\-ting
from the 
duration-hardness relation $E_{\rm p}\propto \tau^{-1/2}$ for
$x=1/4$ (Eq.14). 
\vskip 0.2cm
\noindent
{\it Average Lorentz factor:}\vskip 0.1cm
\noindent
The average Lorentz factor in GRBs closely follows 
the uniform input distribution while large 
values of $\bar \Gamma$ are favored in XRFs.
This is again a consequence of Eq.14 since $E_{\rm p}\propto {\bar \Gamma}
^{-1/2}$ for $x=1/4$.
\vskip 0.2cm
\noindent
{\it Contrast of the Lorentz factor:}\vskip 0.1cm
\noindent
The distribution of the contrast $\kappa$ shows a striking dif\-fe\-ren\-ce
between GRBs and XRFs. XRFs appears to be produced by relativistic winds
where the contrast typically does not exceed a
factor of 4. The maximum of the XRF distribution is located at $\kappa=1.4$. 
Conversely, the proportion of GRBs steadily increases with
$\kappa$. As a consequence of the small contrast of $\Gamma$
in XRFs the 
efficiency
for energy dissipation by internal shocks
\begin{equation}
f={E_{\rm diss}\over mc^2(\Gamma_1+\Gamma_2)}=
{\kappa^{1/2}+\kappa^{-1/2}-2\over \kappa^{1/2}+\kappa^{-1/2}}
\end{equation}
is small, close to 1\% at the maximum of the contrast distribution.
\vskip 0.2cm
\noindent
{\it Injected power:}\vskip 0.1cm
\noindent
The distribution of the injected power $\dot E$ 
shows that large $\dot E$ are favored in both GRBs and XRFs  
because events
with low injected power often escape detection. 
XRFs are therefore not characterized by a deficit of injected 
power (even if the largest $\dot E$ are more frequently found in GRBs).
They appear weak and soft
due to the inefficiency of their internal shocks.
The dissipated energy in XRFs and GRBs is compared in Fig.7. The two
distributions peak at $7\,10^{51}$ and $6\,10^{52}$ erg respectively. 
Again, this difference of nearly a factor of ten 
comes from the lower efficiency of
internal shocks in
XRFs relative to GRBs.
Finally, the shock parameters, through the value of $C_{\rm p}$, also
show some dif\-fe\-rences between GRBs and XRFs. A smaller $C_{\rm p}$ 
naturally favors the production of an XRF but a reduction of $C_{\rm p}$ 
alone is not enough since, contrary to a low $\kappa$, it 
increases the softness without simultaneously decreasing
the ra\-di\-ated power (Barraud et al. 2004b).  
\begin{figure}
\resizebox{\hsize}{!}{\includegraphics{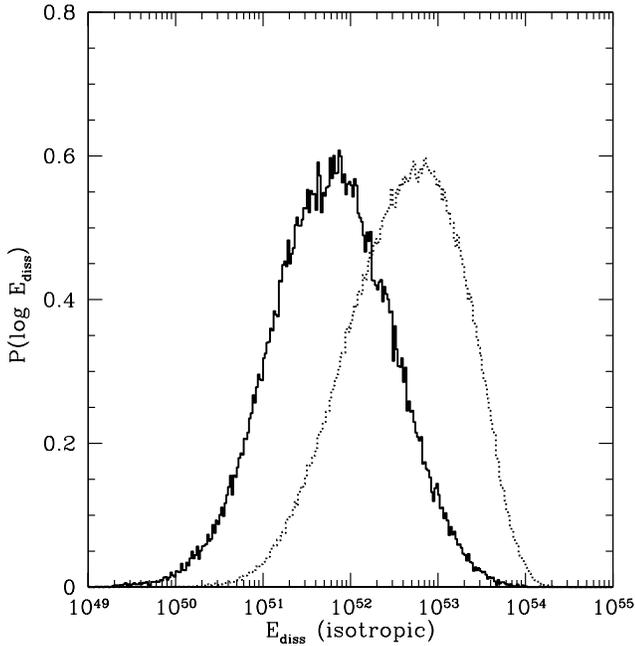}}
\caption{Distribution of the dissipated energy in synthetic 
GRBs (dotted line) and XRFs (full line).}
\end{figure}
\subsection{Amati relation}
Assuming that a constant fraction $\alpha_e$ of the dissipated e\-nergy is
transferred to the electrons and radiated, it is possible to 
check if our synthetic bursts follow the Amati 
relation between the isotropic radiated energy and the value of $E_{\rm p}$ 
in the burst rest frame (Amati et al. 2002). 
The results are shown in Fig.8 for $\alpha_e=0.3$, this rather large value 
of $\alpha_e$ being required to maintain a reasonable overall efficiency 
\begin{equation}
f=\alpha_e\times f_{\rm IS}
\end{equation}
where $f_{\rm IS}$ is the efficiency of dissipation 
by internal shocks.
Bursts which could be detected by HETE 2 have been
represented by large dots in Fig.8. For this sub-group, the best fit by a power law gives
\begin{equation}
E_{\rm p}=200 \left({E_{\rm rad}\over 10^{52}\ {\rm erg}}\right)^{0.46}\ \ {\rm keV}
\end{equation}
\begin{figure}
\resizebox{\hsize}{!}{\includegraphics{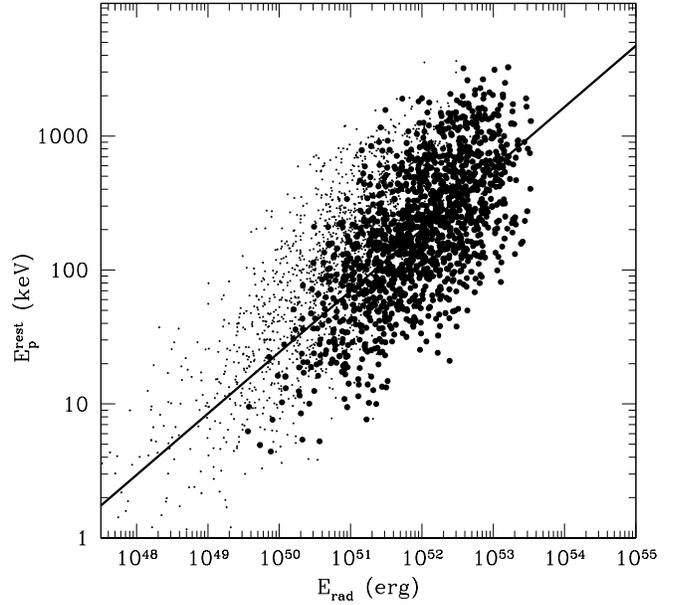}}
\caption{The Amati relation for synthetic burts. A sample of 3000
synthetic events is represented, the large dots correspon\-ding to
those which can be detected by HETE 2. The line is the best fit
for the HETE sub-sample.}
\end{figure}
The exponent is close to 0.5 as found in the observational
Amati relation,
which now extends over five orders of ma\-gni\-tu\-de in $E_{\rm rad}$
when XRF 020903 is included (Sakamoto et al. 2004).   
We however notice that the dispersion of synthetic bursts relative 
to the power law is larger than for observed bursts. If the small 
dispersion of the observational Amati relation 
is confirmed in the future with a larger number of bursts 
it will provide a strong constraint, probably indicating that a new physical
ingredient -- such as a correlation between some of the wind
parameters -- should be included in the models.

\section{Conclusion} \label{conclusion}
We have used a simple internal shock model
to generate a large number of GRBs with different relativistic
wind parameters such as the average Lorentz factor $\bar \Gamma$, the
contrast $\kappa$ between the maximum and mi\-ni\-mum Lorentz factor or the
injected power ${\dot E}$. We adopted a lognormal distribution of the 
intrinsic 
duration and obtained the redshift distribution assuming that the burst 
rate is proportional to the cosmic star formation rate.   
We also assumed standard values for the low and high energy slopes of the
spectrum, $\alpha=-1$ and $\beta=-2.5$ and discussed different possibilities
regarding the shock parameters.
Our aim was to identify the physical conditions
leading to the formation of XRFs. We have found that
our synthetic XRFs exhibit distributions of redshift, duration and injected
power rather similar to those of GRBs but strongly differ in the 
distributions of $\bar \Gamma$ and $\kappa$. XRFs are events where the
contrast of Lorentz factors is small, predominantly between 1 and 2. 

Since we do not know the true distributions of these parameters
we cannot make any prediction on the re\-la\-ti\-ve fraction of XRFs and 
GRBs. With the
uniform distributions adopted for $\bar \Gamma$ and ${\rm Log}\,\kappa$ we
obtain 16\% of XRFs, 27\% of X-ray rich GRBs and 57\% of GRBs. This 
is not in agreement with the HETE results which show approximately
equal fractions of the three classes (Barraud et al. 2004a). If however 
the weight of events with a low contrast of Lorentz factors is increased, for example with 
a distribution
of ${\rm Log}\,\kappa$ which remains uniform but extends to 0.6 only, so that
the maximum contrast is 4 instead of 10, the agreement with the HETE results
is improved with now 31\% of XRFs, 33\% of X-ray rich GRBs 
and 36\% of GRBs. 

If XRFs are produced by internal shocks with a low contrast 
in the Lorentz factor distribution, the efficiency of energy dissipation
is expected to be smaller than in GRBs. This can in principle be tested
from multiwavelength fits of the afterglow lightcurves and spectra.
Such a study including for the first time 
an XRF (XRF 020903) was recently presented
by \cite{lloyd04}. 
The efficiency for XRF 020903 indeed appears to be much smaller 
than what is found 
for the GRBs of the sample. However
all the efficiencies measured by \cite{lloyd04} are 
quite large 
(0.1 for XRF 020903, between 0.4 and 1 for the GRBs) which can be very 
challenging for internal shock models.      

If conversely the Lorentz factor is always highly variable so that
small values of $\kappa$ rarely occur one should
then look for another origin for XRFs such as viewing angle
effects. This possibility has been stu\-di\-ed in detail by 
Yamazaki et al. (2002, 2004).
It supposes that the jets responsible for GRBs are uniform and have sharp 
edges.
If these assumptions are verified, Yamazaki et al. (2002, 2004) have shown that 
many of the 
observed and statistical properties of XRFs can be accounted for 
by assuming a
distribution of the jet opening angle such as $\Delta\theta\propto 
\Delta\theta^{-2}$ (with $10^{-2}<\Delta\theta<10^{-1}$ rad), XRFs being then
obtained for large viewing angles $\theta_v>\Delta\theta$.

Of course, the intrinsic and extrinsic origins for the softness of XRFs do not
exclude each other. A low $E_{\rm p}$ can result from both
a small contrast and a large viewing angle. Their respective 
contributions to the XRF population could be evaluated, for
different distributions
of $\kappa$ and $\Delta\theta$, by adding the viewing 
angle effect
to the present study. 

\begin{acknowledgements}
The authors would like to thank the anonymous re\-feree for an extremely
careful reading of the manuscript. His/her comments and remarks 
have strongly contributed improving our paper. 
\end{acknowledgements}

\end{document}